# Laboratory modelling of solar wind interaction with Lunar Magnetic Anomalies

M. S. Rumenskikh[1], A. A. Chibranov[1], M. A. Efimov[1], A. G. Berezutsky[1], V. G. Posukh[1], Yu. P. Zakharov[1], E. L. Boyarintsev[1], I. B. Miroshnichenko[1,2], P. A. Trushin[1], A. V. Divin[3] and I. F. Shaikhislamov[1]

[1] Laser Physics Department, Institute of Laser Physics, Novosibirsk, Russia
[2] Novosibirsk Technical State University, Novosibirsk, Russia
[3] Physics Department, St. Petersburg State University, St. Petersburg, Russia

E-mail: marina_rumenskikh@mail.ru

**Abstract**

The paper presents results of laboratory experiment modeling the interaction between Lunar magnetic anomalies and Solar wind. To model the LMA we use quadrupole magnetic field. The main dimentionless parameter of the problem, the ion inertia length relative to the mini-magnetosphere size, well corresponds between experiment and LMA conditions. The main result is measurement of the magnetically reflected proton fluxes, which show qualitative agreement to available satellite data.

## 1. Introduction

The processes occurring in space plasmas are complex and difficult for interpretation. Their study in situ and in the laboratory is an important task. The plasma environment of the Moon requires detailed understanding for future explorations. Data obtained by satellite measurements (*Wieser et al 2010; Richmond et al 2008; Halekas et al 2001; Tsunakawa et al 2015; Halekas et al 2008; Lin et al 1998*) showed that Moon does not have a global magnetic field. When hitting the Lunar surface, which is not protected by a magnetic field, protons of the Solar Wind (SW) recombine with electrons, and the fluxes reflected from the surface are mainly represented by neutral particles. Solar particles also charge the lunar surface (*Stubbs et al, 2005*). However, solar protons are reflected in a qualitatively different way in several places on the Moon. There are considerable fluxes of the reflected protons while backscattered neutral particle flux is reduced above such areas (*Skalsky et al 2015; Saito et al 2008; Saito et al 2012; Wieser et al 2009; Holmstrom et al 2010*). This phenomenon is explained by the existence of the areas of local magnetization of the lunar crust, which are called Lunar Magnetic Anomalies (LMA). Local magnetic fields form mini-magnetospheres around such areas partially preventing direct penetration of charged particles to the surface, which results in changes of the local albedo of the Lunar crust (*Lin et al 1998*).

It is known that LMAs are located mainly over places with uneven surface structure (craters and oceans). The theoretical study predicted and measurements of Lunar Prospector revealed the existence of "Lunar swirls" in the areas of LMA with affected local albedo of the lunar surface (*Hood et al 1980; Garrick-Bethell et 2011; Blewett et al 2011; Denevi et al 2016; Deca et al, 2020*). Hemingway and Garrick-Bethell (2012) have inferred that the dark portions of the patterns, typical of old regolith, correspond to cuspal features of magnetic fields, while the bright portions, surmised to consist of young regolith, correspond to the magnetic domes. *Gattacecca et al 2010* studied magnetic features of lunar rocks and described two mechanisms – thermoremanent and shock remanent magnetization.

Detailed studies of lunar mini-magnetospheres are presented mainly by numerical methods (*Holmstrom et al 2010; Kallio et al 2012; Deca et al 2014; Deca et al 2015*). Laboratory modeling of these structures is promising since the main similarity parameters are much easier to perform for a mini-magnetosphere than for the earth-type magnetospheres. In the laboratory experiment described in (*Bamford et al. 2012*), a plasma jet impacted a magnet at conditions when ion gyroradius and ion inertia plasma length significantly exceeded the size of the magnetic field localization region. Measurements showed the stopping of ions and the formation of electrostatic potential on the scale of the electron inertia plasma length. In the report of *Jia Han 2017* experiments are described on the interaction of dipole magnetic field with the nitrogen ion beam. It was shown that increasing ion energy decreases the surface potential so the shielding becomes less efficient.

In our previous works (*Shaikhislamov et al. 2013, 2014; 2015a; 2015b*), a mini-magnetosphere was studied in Terrella type laboratory experiments, 2.5D numerical simulation by the Hall MHD code, and theoretical analysis. In particular, experiments have shown that when the value of the ion inertial length $L_{pi} = c/\omega_{pi}$ is greater than the distance to the balance point $R_p$ of kinetic and magnetic pressures, $B^2(R_p)/8\pi = m_p n v^2$, the plasma penetrates into the magnetosphere and stops at the Stoermer's radius of the minimum convergence of the ions. Another feature is that as the $L_{pi}$ increases, the measured magnetopause position $R_m$ shifts further away from the pressure balance distance $R_b$, and the field jump on the magnetopause decreases. Based on experimental data, a complex model was constructed for the first time, which explains the most important features of the mini-magnetosphere observed in (*Shaikhislamov et al. 2013, 2014*) and in other works based on numerical simulations (*Fujita 2004; Omidi et al. 2002*).

Complex spatial LMA fields might be represented as a superposition of irregular placed dipoles (*Kurata et al. 2005, Purucker and Nicholas 2010, Ravat 2020*). In our recent experiment on laboratory modeling of LMA and its interaction with plasma flow we have employed and compared dipole and quadrupole configurations of the magnetic field (*Rumenskikh et al. 2020)*. Also, we measured for the first time in laboratory experiments on mini-magnetosphere the magnetically reflected ions. It was found out that up to 50% of incoming ions are reflected by magnetic field and that reflected flux has a complex spatial structure. These features are qualitatively similar to satellite measurements above LMA.

As is argued in (*Shaikhislamov et al. 2013, 2014, 2015*), one of the crucial parameters which determine the interaction of LMA with solar wind is Hall parameter $D=R_b/L_{pi}$. In *Rumenskikh et al. 2020* it was larger than unity $D \approx 1 \div 1.5$. Considering conditions at Moon, it is supposed that at strongest LMAs there exists an area of about ~10 km in height partially protected from the direct penetration of the SW plasma, while the estimated distance to the pressure balance point between SW pressure and magnetic field of LMAs is about ~30 km (*Wieser et al. 2010; Wang et al. 2012; Lue et al. 2011*). With the ion inertia length of the SW being ~100 km, the Hall parameter above LMA is substantially smaller than unity, $D\sim0.3$. In the present work, we explored the conditions more similar to the interaction between solar wind and LMA with the Hall parameter $D\sim0.5$. This is a principal novelty and difference from *Rumenskikh et al. 2020*, which allowed us to see new effects important to the subject of the plasma environment above LMAs.

The laboratory experiment presented in this article is notable for a number of reasons. First, two different regimes of plasma velocity and density were investigated. This is important because of the variability of space weather conditions and solar activity. Secondly, by using a quadrupole source we take into account the complex nature of LMA fields, which are characterized by a more spatially compact structure than the simple dipole configuration. The third novelty of our experiments is the combination of Faraday collector, electric probe, and magnetic probe diagnostics, which greatly complement each other and allow comparing results with data of satellites on the moon orbit. In particular, we compare the flux of magnetically reflected ions, the interaction characteristic which attracted a lot of attention in recent satellite missions to the Moon, such as SELENE Explorer (*Saito et al. 2010*), Chandrayaan-1 (*Wieser et al. 2010*), Chang'E-2 (*Wang et al. 2012*).

## 2. Experimental setup

The KI-1 Facility was designed to study a variety of processes occurring in space plasma. The vacuum chamber (Fig. 1) is equipped with an induction discharge theta-pinch plasma source to generate quasi-homogeneous and quasi-stationary plasma flows. The volume of the pre-ionizer is filled with a gas, in our case molecular hydrogen, and the induction electric field created by the pulsed magnetic field of the solenoid induces a breakdown and ionization of the gas, as well as compression and heating of the plasma. The generated plasma is injected in the vacuum with residual pressure ~ $10^{-6}$ Torr. To stabilize the plasma flow in the direction of propagation the uniform magnetic field is applied along the chamber axis. In present experiments it was small, 5 G, and unimportant for the processes under consideration.

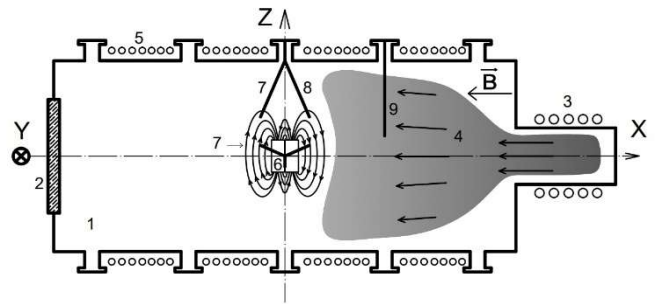

Fig. 1. Layout of the experiment: (1) 5 × 1.2-m vacuum chamber, (2) observation window, (3) theta pinch, (4) hydrogen plasma flow, (5) solenoid, (6) magnetic quadrupole, (7) conjugated magnetic and electric probes, and (8, 9) collectors.

The source of the magnetic field was made of two compact dipoles, each consisting of multi-coil put in epoxy shell 5 cm in diameter. The coils were powered in parallel by a current pulse with a duration of ~ 200 μs, and each could generate a magnetic moment μ of up to $10^6$ G·cm$^3$. When the dipole moments of coils were switched in parallel, a dipole magnetic field was created, and when they were switched in antiparallel a quadrupole field was generated. The magnetic moments of the coils were directed along the Z axis, and their centers were spaced along the X axis at a distance of Δ = 2.5 cm for one and -2.5 cm for the other. Such geometry is the simplest for studying the interaction with the plasma flow, since in this case the zeros of the field are located in the y−z plane. Along the direction of the plasma flow, i.e. the X axis, the field value for parallel switching of coils (dipole) or antiparallel (quadrupole) is equal to B = $B_z = \frac{\mu}{(x-\Delta)^3} \pm \frac{\mu}{(x+\Delta)^3}$. We note that if for a dipole the field decreases with distance as $r^3$, for a quadrupole it decreases faster, $\sim r^4$.

The diagnostics used were ion collectors, magnetic and electric probes. Measurements of particle fluxes were made with two collectors (termed further on as KB1 and KB2) located at a distance of 10 and 70 cm from the quadrupole center. The collecting electrode was inside of the 1.5 mm guide tube with 0.6 mm input hole. The guide tube gave translational and rotational degrees of freedom. The collector farthest from the source of the magnetic field moved along the Y axis, with other coordinates fixed at Z = 0, X = 70 cm. The nearest collector was positioned at an angle of 45º to the Y-Z plane with the closest distance to the quadrupole center equal to X=10 cm. The accuracy of the spatial positioning of collectors was 0.5 cm. The guide tube could rotate around its axis, which allowed detecting particle flows from different directions determined by the orientation of the input hole. Further on, the orientation angle α=0º corresponds to measuring the flux from the theta-pinch, while the α=180º to measuring reflected ions. The angular resolution diagram of collectors was about ±25º.

Magnetic probes consisted of three miniature orthogonally arranged coils with a total area of about 1 cm² each. The time resolution of the probes was 20 ns, the bandwidth was 10 MHz, and the spatial resolution was 0.5 cm. The device for moving the probes allowed covering the area ±25 cm from the center of the dipole. Langmuir's electric probes measured the ion current and floating potential of the plasma. The calculation of the ion concentration from the measured current based on calibration data is described in *Shaikhislamov et al. 2015b*.

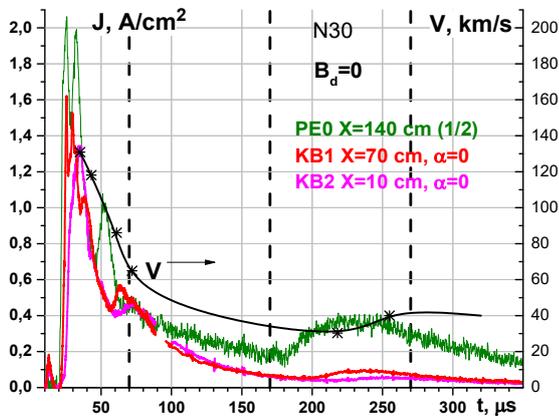

Figure 2. Oscilloscope signals of the plasma ion current at different distances from the quadrupole (left axis), measured by an electric probe (PE0, the signal is reduced by half) and collectors (KB1 and KB2, both oriented in the direction of flow) in the absence of a magnetic field. The black curve shows the velocity of plasma in the quadrupole region (right axis). The zero moment of time corresponds to the discharge of theta pinch. Vertical lines mark the time intervals for averaging signals that are used in further analysis.

The theta-pinch plasma has a dynamic character and a shot-to-shot variability. The oscilloscope signals presented in Fig. 2 show three distinct phases in time. The first flow until 70 μs was the fastest and shortest, and had too high plasma density for the purposes of this work. The second flow in the interval 70-170 μs and the third one in the interval 170-270 μs were used for the data analysis. The average plasma concentration in the region close to the quadrupole in the specified time intervals was $1.6 \times 10^{12}$ and $4.6 \times 10^{11}$ cm$^{-3}$ respectively. The plasma velocity calculated with the time-of-flight method was about 30-60 and 30-40 km/s in the first and second intervals respectively (Fig. 2). As will be shown below, the variations in plasma density and velocity affected the interaction with the magnetic field.

| Interaction regime | n, cm$^{-3}$ | V, km/s | $\mu_B$, G×cm³ | $L_{pi}$, cm | $R_b$, cm | D | Kn |
|---|---|---|---|---|---|---|---|
| High pressure | $4.5 \times 10^{11}$ | 60 | $1.7 \times 10^5$ | 33,9 | 23,5 | 0,69 | 31 |
| Low pressure | $2 \times 10^{11}$ | 30 | $1.7 \times 10^5$ | 50,8 | 33,9 | 0,66 | 3 |
| Lunar mini-magnetosphere | 10 | 300 | $\sim 1,8 \times 10^{12}$ | $\sim 10^7$ | $3 \times 10^6$ | 0,3 | $\sim 10^7$ |

Table 1. Dimension and dimensionless parameters of the laboratory experiment and typical lunar mini-magnetosphere. Column designations (from left to right): ion concentration, average velocity, magnetic moment, ion plasma length, distance to the pressure balance point, Hall parameter, Knudsen number.

## 3. Results

Our previous work *Rumenskikh et al. 2020* has shown that the interaction between quadrupole magnetic field with moment μ=$5 \times 10^5$ G*cm³ and plasma flow with a density of $\sim 5 \times 10^{11}$ см$^{-3}$ resulted in creating a mini-magnetosphere with a size of about 20 cm. This process was accompanied by significant disturbance of the magnetic field, as evidenced by an increase of the main component of the magnetic field $B_z$. Despite plasma penetration inside the magnetosphere, a significant decrease in plasma density was observed at the distances <15 cm from quadrupole.

The interaction regime of the present experiment significantly differed from the previous one by having lower plasma density and weaker magnetic field. Fig. 3 shows measurements by probe located close to the quadrupole in the time interval corresponding to the first, fastest and densest, plasma flow. The reduction of ion density inside the mini-magnetosphere is small, even as close to the magnetic field source as 7 cm. Moreover, magnetic disturbances of quadrupole field caused by plasma were less in amplitude by several times than in the previous experiment, and they couldn't be distinguished against the chaotic fields generated in theta-pinch and carried by plasma.

The direct evidence of interaction with magnetic field gives the floating plasma potential. The behavior of plasma potential in magnetic field has been analyzed in our previous papers (*Shaikhislamov et al. 2009, 2014, 2015b*). In the absence of the quadrupole field the floating potential had a negative sign because it is produces by electrons. In presence of magnetic field it shows a positive signal, which points to the blocking of electrons by the magnetic field. We note that the maximum of the potential approximately corresponded to the energy of the fastest ions. Also, in the present experiment, like in our previous experiments with the dipole magnetic field (*Shaikhislamov et al. 2014*), high-frequency oscillations of the magnetic field were observed with the frequency of about ω ≈ $3*10^7$ s$^{-1}$ (Fig. 4). This value well corresponds to a low-hybrid frequency $\omega_{cei}$ at the magnitude of magnetic field of about 150 G (at the probe location of X=7 cm).

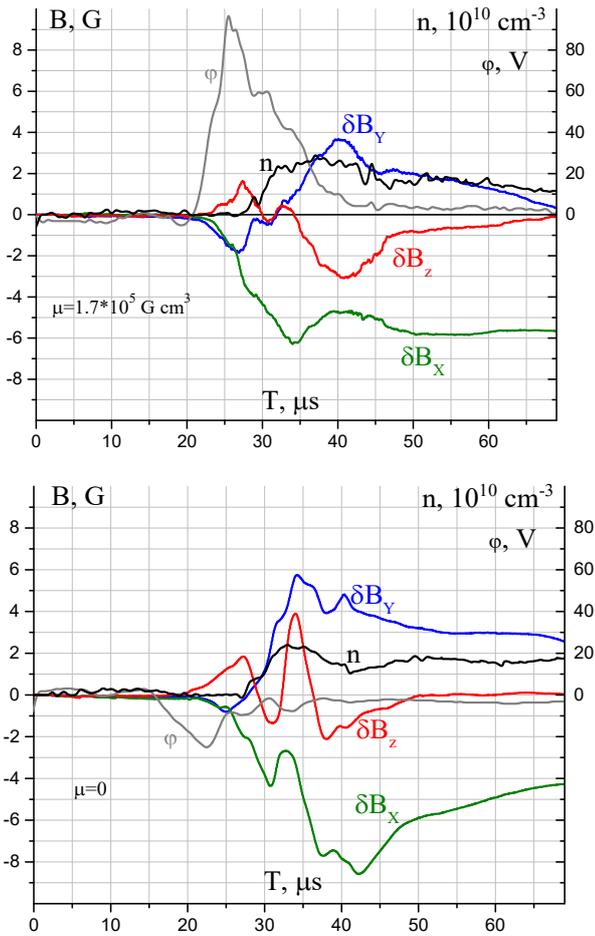

Fig. 3. Magnetic field variations ($\delta B_Z$-red; $\delta B_X$-green; $\delta B_Y$-blue), plasma density (n-black) and floating plasma potential (φ-grey) measured by probe in the quadrupole field (left) and in the absence of magnetic field (right). The probe was located at X=7 cm, Y=0, Z=0.

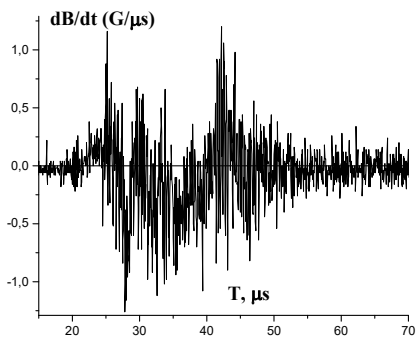

Fig. 4. Derivative of the magnetic field measured by probe in the plasma flow in a quadrupole field.

The main experimental novelty of the present work is measurement of ion fluxes. It was made by ion collectors located at distances of 10 and 70 cm respectively. They detected the fluxes of ions coming from the source of plasma and reflected back by the quadrupole magnetic field. Fig. 5 demonstrates the oscilloscope signals of Langmuir probes and collectors with and without magnetic field. The comparison reveals a significant effect of the quadrupole field on the backscattered ions. When magnetic field of the coils is switched off the collectors show comparatively weak flux of ions captured by collectors due to finite resolution of the directional pattern and the thermal spread of ion velocities. Switching on the magnetic field increases the reflected flux by 5 times at a distance of X=10 cm from the quadrupole center (bottom panel of Fig. 5). At a much larger distance of X=70 cm the effect is also observed, but not so significant (increase by a factor of 1.5, upper panel of Fig. 5).

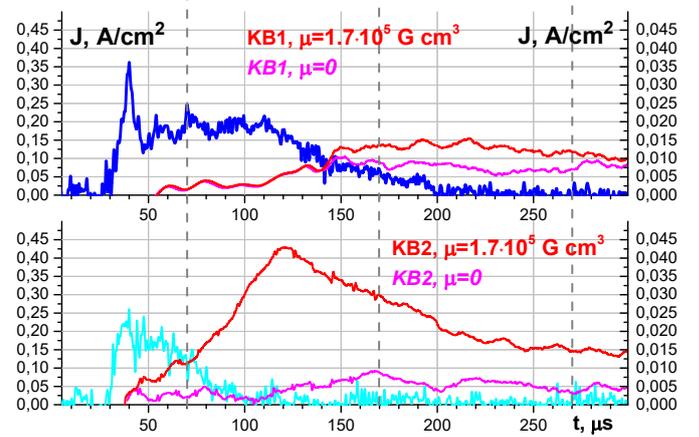

Fig. 5 Oscilloscope signals in magnetic field and without it at different distances from the quadrupole. The blue and cyan lines show the current of Langmuir probe with and without magnetic field respectively. The probe is located in close proximity to the center of the quadrupole, X=7 cm, the values are given at the left axis. The red and magenta curves show the flux measured by collectors with and without magnetic field respectively, the values are given at the right axis. The top panel shows signals of remote collector placed at a distance of X=70 cm, while the bottom panel – signals of close-in collector placed at a distance of X=10 cm. The collectors are oriented to the quadrupole (α=180°) and measure the reflected flux. The vertical dotted lines mark the intervals for time averaging of signals used in further data representation.

In addition, the Langmuir probe placed close to the quadrupole center at the point X=7 cm shows almost complete blocking of the incoming flow by the magnetic field starting from the time ~110 μs (bottom panel). This is direct evidence that in the realized experiment there was an area near the quadrupole that was completely protected from the direct ion penetration. Such an effect is observed over strong LMAs.

In addition to the rotational degree, the collectors had a translational degree of freedom along the guide tube. For the KB1 collector, the direction of motion coincided with Z axis. For the KB2, the line of movement was at the angle 45° to Z and Y axes.

For the quantitative analysis of the reflected fluxes the collector signals were averaged in the chosen time intervals as was specified above and shown in Fig. 2 and 5, $\langle J \rangle = T^{-1} \int J dt$.

To reveal the particular effect of the magnetic field, the particle flux measured without the magnetic field was subtracted: $\langle J \rangle_B - \langle J \rangle_{B=0}$. To calculate the fraction of reflected particles, this quantity was normalized to the incoming ion flux: $F = (\langle J \rangle_B - \langle J \rangle_{B=0}) / \langle J \rangle_{\alpha=0, B=0}$.

The spatial distribution of the reflected ions measured nearby and far from the quadrupole is presented in Fig. 6. The L-axis origin is at the equatorial plane XY. As can be seen the collector at X=10 cm measured the reflected flux with magnitude up to 50% of the incoming flux. At the same time, there were spatial variations between the regions below and above the equatorial plane, and temporal variations between first and second averaging intervals related to denser and more rarified plasma. In the far zone, the reflected flux was smaller by about an order of magnitude. Moreover, the reflection during the first time interval didn't exceed the measurement error.

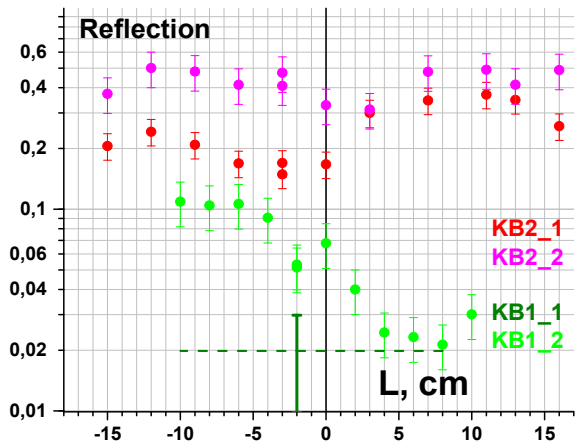

Fig. 6. Spatial distribution of the fraction of reflected ions along the line of collectors' motion. Data for collector KB2 are shown by red symbols for the time interval 70-170 μs and by magenta for the time interval 170-270 μs. Green symbols show data for collector KB1 for the time interval 170-270 μs. The dotted olive curve shows the average level of reflection measured by KB1 in the interval of 70-170 μs.

## 4. Discussion and Conclusion

Lunar magnetic anomalies are characterized by complex multipole structure of field and fluxes of magnetically reflected protons of SW are directly related to this. In the present work this feature has been studied in laboratory model of LMA for the first time. Taking into account a comparatively small size of the LMAs, the main values characterizing the interactions are the distance $R_b$ from the crust (field source) at which the kinetic pressure of plasma flow is balanced by magnetic pressure, and the ion inertia length $L_{pi}$. For the strongest LMA $R_b$~30 km and the typical value of $L_{pi}$ is equal to 100 km. The Lunar Prospector has detected the cases of distinct decreasing of proton density over the strongest LMA approximately at this altitude of ~30 km (Wieser et al. 2010; Wang et al. 2012; Lue et al. 2011). Thus, the Hall parameter for LMA is significantly smaller than unity, $D=R_b/L_{pi}$~0.3. In the realized experiment, this characteristic parameter has values close enough to those for the LMA, D<1. A region close to the quadrupole (X<=7 cm) was observed in the experiment which was completely blocked by the magnetic field from incoming protons, the same as for LMA.

Let's compare magnetic fields inducted by plasma flow around the source of the magnetic field. The characteristic value of such fields is determined by the pressure of the upstream plasma $B_o = \sqrt{8\pi m n_o V_o^2}$, which is equal to 50 nT in case of SW at Earth orbit. Unlike the Earth magnetosphere where induced fields of such magnitude are present, they are not observed over LMAs. The lack of magnetic field compression by plasma flow is obviously associated with such property of the mini-magnetosphere as the absence of a bowshock. In our previous experiment (Rumenskikh et al. 2020) a transitional regime of interaction has been observed with the induced field at a level approximately equal to the supposed magnitude $\delta B = \sqrt{8\pi m_p n_o V_o^2} \approx 25$ G. In the present experiment this value should be about 11 G for the low pressure mode. If the magnetic variation of such value was present, then it would be measured. However, as it is shown in Fig. 3 the induced variations were lower than 3 G. Thus, the present laboratory simulation of mini-magnetosphere is principally different from our previous work Rumenskikh et al. 2020 and much better corresponds to the properties of LMA's plasma environment.

The main result of this work is measuring the fluxes of magnetically reflected ions. The direct comparison of measurements inside of a mini-magnetosphere (which were obtained in the experiment) with the data for the Moon is currently unavailable because the orbits of satellites with rare exceptions lie higher than the areas above the LMA shielded from SW protons. However, such a comparison is possible for the flux of reflected protons. At distances close to the quadrupole, which for the LMA translate to heights ~15 km, the reflected flux was up to 50% (with an average value of 40%) of the incoming flux of protons. Thereby, a significant part of protons not penetrating into the magnetically protected area are scattered back. At the same time, the normalized reflected flux was appreciably larger for the second time interval with lower density.

At the distance of ~70 cm the reflected flux was significantly lower, 2-10%, with an average value of about 6%. Herewith the reflected protons were detected only in the second time interval with low plasma density. In the first time interval with higher density, the value of reflected flux didn't exceed the measurement error. In the present experiment, a distance of 70 cm corresponded in units of ion plasma length to 140 km of the altitude above the LMA. This is slightly higher than the typical orbits of satellites ~100 km, which measured above the strongest LMA magnetically reflected fluxes up to ~20% of the SW flux (Saito et al. 2010). At lower orbits, the fraction of reflected flow could reach up to 50% (Lue et al. 2011). The significant reduction of reflected flux with distance is explained by the finite size of the reflection area. As was described, the electric probe located at the distance of 7 cm from the center of quadrupole shows the total blocking of plasma flow. The collector located at a distance of 10 cm shows the reduction of the incoming flux by about 2 times (in measurements with α=0° μ=0 and μ=1.7·10⁵ G·cm³). At the distances of X>16 cm the electric probe shows no perceptible effect of the magnetic field on plasma flow. Thereby, the size of region of significant proton reflection has a radius of about 10-15 cm. Outside this area, the reflected flux falls as $r^{-2}$, so at the distance of 70 cm from the quadrupole, the attenuation could reach 20 times. The observed attenuation in experiment was about 7 times, which points to a more complex geometry of the reflection region and fluxes.

Significant and sharp modulations of the reflected fluxes are typical of LMA (*Saito et al. 2008*). In the present and our previous experiment reflected fluxes had a complex behavior also, showing spatial modulation by several times (for example, variation from 2% to 10 % of KB1 and KB2 data in Fig. 5). Such a difference observed in the experiment is explained, most likely, by the Larmor rotation of protons, which creates an asymmetry along the Y-axis. Since the far collector moved along both the Z and Y axes, this asymmetry in measurements is to be expected. The measurements showed that in the considered time intervals (t > 70 μs) electric potential of plasma had a negative value and was about ~1 V of magnitude. Therefore, the reflection of ions is not associated with electric field of charge separation and occurs due to the Larmor rotation. Such a gyration process of reflected ions in case of mini-magnetosphere was demonstrated in the laboratory experiments and numerical simulations (*Shaikhislamov et al. 2013; 2015a*). Similarly to the dipole configuration of the magnetic field, one can estimate the distance of minimum convergence of ions (the Stoermer radius) for a quadrupole configuration by the formula:

$$\int_{R_{st}}^{\infty} \frac{eB_z}{m_p c} dx = V_o$$

For the measured velocity of 40 km/s we obtain $R_{St} \approx 13$ cm, and this value corresponds well to the obtained measurements.

The realized experiment reproduced the mini-magnetospheres above LMA qualitatively and to a good extent quantitatively by the main dimentionless parameters and the structure of the magnetic field. Measured values of reflected flux correspond to the available satellite observations. The agreement between laboratory and satellite measurements validates further research in this field. In particular, the direct comparison with numerical kinetic modeling for verification of existing codes can be of great importance (*Deca et al. 2016*). In further experiments, it is possible to reproduce more accurately the structure of magnetic fields, as well as partial reflection and recombination of protons on the dielectric surface that models the Moon's crust. Another aspect for modeling is the direction of the SW relative to the crust, including the interaction on the limb of the Moon.

## Acknowledgements

The work was supported by the RFBR projects 19-02-00993 and 18-29-21018, as well as within the project of the Ministry of science and higher education of the Russian Federation (№ AAAA-A17-117021750017-0).